\documentclass[useAMS,usenatbib]{mn2e}
\usepackage{amsmath,fleqn,graphicx,amssymb}
\usepackage{multirow}
\def\bolth{\mbox{\boldmath$\theta$}}

\def\kpc{\,\mathrm{kpc}}
\def\kms{\,\mathrm{km\,s}^{-1}}

\def\bolv{\mathbf{v}}

\def\rmd{\,\mathrm{d}}
\def\kpc{\,\mathrm{kpc}}
\def\kms{\,\mathrm{km\,s}^{-1}}
\def\vsol{\mathbf{v_\odot}}
\def\vsfr{\mathbf{v}_\mathrm{SFR}}
\def\vpsfr{v_{\phi,\mathrm{SFR}}}
\def\fixed{\mathrm{fixed}}
\def\vary{\mathrm{\ free}}
\def\data{\mathrm{data}}
\def\vmas{v_\mathrm{m}}
\def\HMSFR{{\sc hmsfr}}
\title[Uncertainty in Galactic parameters]
{The uncertainty in Galactic parameters}

\author[P.~J.~McMillan \& J.~J.~Binney]{
  Paul~J.~McMillan\thanks{E-mail: p.mcmillan1@physics.ox.ac.uk},
  and James~J.~Binney \\
  Rudolf Peierls Centre for Theoretical Physics, 1 Keble Road,
  Oxford, OX1 3NP, UK
}

\begin{document}
\maketitle

\begin{abstract}
  We reanalyse the measurements of parallax, proper motion, and
  line-of-sight velocity for 18 masers in high mass star-forming
  regions presented by~\cite{Reidetal2009}.  We use a likelihood
  analysis to investigate the distance of the Sun from the Galactic
  centre, $R_0$, the rotational speed of the local standard of rest,
  $v_0$, and the peculiar velocity of the Sun, $\vsol$, for various
  models of the rotation curve, and models which allow for a
  typical peculiar motion of the high mass star-forming regions.

  We find that these data are best fit by models with non-standard
  values for $\vsol$ or a net peculiar motion of the high mass
  star-forming regions. We argue that a correction to $\vsol$ is much more
  likely, and that these data support the conclusion
  of~\cite{Binney2009} that $V_\odot$ should be revised upwards from
  $5.2\kms$ to $11\kms$. We find that the values of $R_0$ and $v_0$
  that we determine are heavily dependent on the model we use for the
  rotation curve, with model-dependent estimates of $R_0$ ranging from
  $6.7\pm0.5\kpc$ to $8.9\pm0.9\kpc$, and those of $v_0$ ranging from
  $200\pm20\kms$ to $279\pm33\kms$. We argue that these data cannot
  be thought of as implying any particular values of $R_0$ or $v_0$.  
  However, we find that $v_0/R_0$ is better
  constrained, lying in the range $29.9-31.6\kms\,\kpc^{-1}$ for all
  models but one.

\end{abstract}


\section{Introduction}\label{sec:intro}
The fundamental parameters that define the Solar position and
velocity within the Galaxy remain uncertain to a remarkable
degree. The major remaining uncertainty is that in the distance from
the Sun to the Galactic centre, $R_0$. The most recent results from
studies of stellar orbits in the Galactic centre
\citep{Ghezetal2008,Gillessenetal2009} give values of $R_0 =
(8.4\pm0.4)\kpc$ and $R_0 = (8.33\pm0.35)\kpc$ respectively.  These
can be compared to the earlier estimate in the review
by~\cite{Reid1993} of $R_0 = (8.0\pm0.5)\kpc$.

The total velocity of the Sun about the Galactic centre is the sum of
the velocity of the local standard of rest (LSR), $v_0$, and the
peculiar motion of the Sun with respect to the LSR in the same
direction, $V_\odot$. This total velocity can be determined using the
apparent proper motion of Sgr A*, $\mu_{A*}$, since it is expected to
be moving with a peculiar velocity less than $\sim1\kms$ at the
Galactic Centre. This constraint on the peculiar motion of Sgr A* is
justified by the observation that the velocity of Sgr A* perpendicular
to the plane is consistent with zero, with uncertainties $\sim1\kms$
\citep{ReidBrunthaler2004}, and because this motion is thought to be
due to stochastic forces from discrete interactions with individual
stars, so the velocity in the plane should be similar to that
perpendicular to it \citep*{Chatterjeeetal2002}.
\cite{ReidBrunthaler2004} found
$\mu_{A*}=(6.379\pm0.024)\,\mathrm{mas\ yr}^{-1}$, which corresponds
to $(v_0+V_\odot)/R_0 =(30.2\pm0.2)\kms\kpc^{-1}$, or a velocity about
the Galactic centre (using the \citeauthor{Ghezetal2008} result for $R_0$)
of $v_0+V_\odot = (252\pm11)\kms$, where by far the dominant
uncertainty comes from the value of $R_0$. Analysis of the GD-1
stellar stream \citep*{Koposovetal2009} has recently been used to
suggest a significantly lower value, $v_0=221^{+16}_{-20}\kms$.

The velocity of the Sun with respect to the local standard of rest,
$\vsol$, is often assumed to be well known and constrained to within
$\sim0.5\kms$ because of the analysis of the dynamics of nearby stars
conducted by~\citet[henceforth DB98]{DehnenBinney1998b}, and again
more recently, using identical techniques,
by~\cite{AumerBinney2009}. DB98 found
\begin{equation}
  \begin{array}{ll}
    \vsol &\equiv (U_\odot,V_\odot,W_\odot) \\
    &= (10.00\pm0.36,\,5.25\pm0.62,\,7.17\pm0.38)\kms,\\
  \end{array}
\end{equation}
which has been widely accepted and used.  However~\citet[henceforth
B09]{Binney2009} suggests that the value for $V_\odot$ determined in
these papers may be an underestimate by $\sim6\kms$. This is because
the analysis by DB98 uses Stromberg's equation, which is derived under
the assumption that the Galactic potential is axisymmetric, and
extrapolates to zero velocity dispersion.  In practice, the Galactic
potential is not axisymmetric, and the smaller the velocity dispersion
of a population, the more it is affected by departures from
axisymmetry. B09 uses a more global approach, ensuring that more
emphasis is placed on stellar populations with high velocity
dispersions, which one would expect to be less affected by the
non-axisymmetry of the Galactic potential.

\cite{Reidetal2009} brought together observations of masers seen in
high mass star-forming regions (\HMSFR s) in the Milky Way. 
They used simple statistical tools in an effort to determine the
values of $R_0$ and $v_0$, using the DB98 value of $\vsol$, initially
under the assumption that the \HMSFR s were moving on circular orbits
in a flat rotation curve, for which they offer best-fitting parameters
$R_0 = (8.24\pm 0.55)\kpc$ and $v_0 =(265\pm 26)\kms$, but with a high
$\chi^2$ value.

In addition they considered a model in which the \HMSFR s were moving
with a characteristic velocity with respect to their local circular
velocity. They found that they achieved a significantly improved fit
to their data using this model with a peculiar velocity $\sim 15\kms$
in the opposite direction to rotation. This fit yielded $R_0 =
(8.40\pm 0.36)\kpc$ and $v_0=(254\pm16)\kms$. They briefly considered
a model in which the value of $\vsol$ was allowed to vary, finding an
acceptable fit with $\vsol=(9,20,10)\kms$. However, believing the DB98
result to be ``well determined'' they did not pursue the matter
further.

In this paper we re-examine the data described in \cite{Reidetal2009},
and conduct a likelihood analysis for various models of the velocity
distribution of the maser sources. This enables us to exploit these
data more thoroughly. We also consider the implications of the results
from B09 on the interpretation of these data. In
Section~\ref{sec:methods} we explain what the data consist of,
describe our models and our statistical technique; in
Section~\ref{sec:results} we give the raw results and examine their
significance. We discuss the implications of these results in
Section~\ref{sec:discuss}.

\section{Methods}\label{sec:methods} \subsection{The data} These data, as
given in Table 1 of \cite{Reidetal2009}, consist of measurements for
18 masers of: Galactic coordinates ($l_i$, $b_i$), which we can assume
to be exact; parallaxes $\pi_i$; proper motions $\mu_{x,i}$ and
$\mu_{y,i}$; line-of-sight velocities $v_{\mathrm{LSR},i}$. For each
quantity Reid et al.\ give an error and we assume that this error
together with the measured value of the quantity defines a Gaussian
probability distribution for the true value of the quantity.  For the
proper motions $\mu_x$ and $\mu_y$ we shall use the more conventional
notation $\mu_\alpha$ and $\mu_\delta$. The line-of-sight velocity
$v_{\rm LSR}$ is relative to an obsolete estimate of the LSR. 
Fortunately the underlying heliocentric
line-of-sight velocity $v_r$ can be recovered from $v_{\rm LSR}$
without impact on the associated uncertainty
\citep[][Appendix]{Reidetal2009}.
 
\subsection{Our models}

The motion of both the Sun and the masers is dominated by circular
motion around the Galactic centre with velocity
\begin{equation}
  -v_c(R)\mathbf{e_\phi},
\end{equation}
where $v_c(R)>0$ and the minus sign reflects the fact that the Galaxy
rotates clockwise in our coordinate system. We explore three forms for
$v_c(R)$
\begin{itemize}
\item A flat rotation curve, $v_c(R) = v_0$, with $v_0$ being a
  (positive) free parameter.
\item A power-law rotation curve $v_c(R) = v_0(R/R_0)^\alpha$, with
  $v_0$, $R_0$ and $\alpha$ being free parameters.
\item A rotation curve corresponding to that given by the Galactic
  potential Model I in \S 2.7 of~\citet[henceforth GDII]{GDII},
  linearly scaled to variable values of $R_0$ and $v_0$.
\end{itemize}
In each case, we take the parameters of $v_c(R)$ to have uniform prior
probability distributions. This choice ensures that we determine what
\emph{these} data tell us, without prejudice.

\begin{figure}
  \centerline{\resizebox{\hsize}{!}{\includegraphics{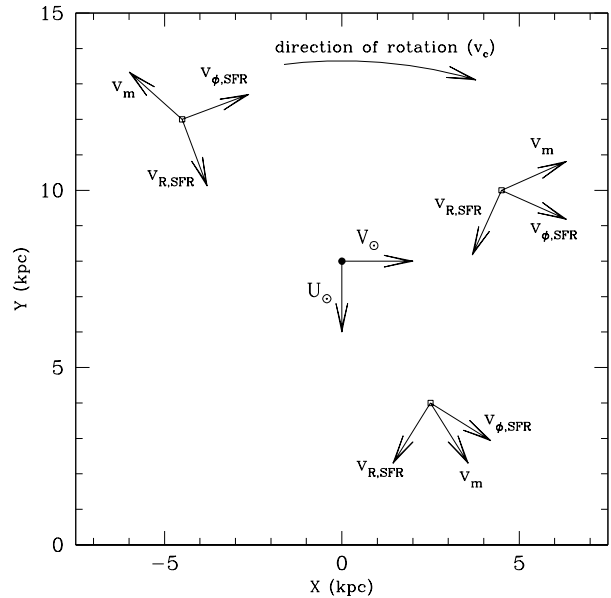}}}
  \caption{Diagram showing a pole-on view of the Galaxy 
    illustrating the various peculiar velocities we consider 
    (e.g. equation~\ref{eq:vbar}). This therefore only shows the in plane 
    components. The Sun (represented by a solid circle) is placed at 
    $(0,8)\kpc$, and components of its velocity $\vsol$ are indicated. 
    Three points (empty squares) are plotted to represent masers, and 
    the directions of the components of any systematic offset from 
    circular velocity in each case ($\vsfr$) are shown, as is the 
    peculiar velocity due to any bias in the observed radial velocity, 
    $\vmas$. This diagram is purely illustrative and does not show any real 
    data.
\label{fig:whatvels}
}
\end{figure}

The velocity of a maser can be expected to differ from the local
circular speed. We separate this difference into a random component
and (in some cases) two systematic components. Like
\cite{Reidetal2009}, we consider the possibility that the velocity of
\HMSFR s has a systematic offset from the circular velocity,
$\vsfr$. We also consider the possibility that the expansion of the
shell within which a maser occurs will displace the maser's velocity
from that of the star, and if we are biased towards seeing masers on
the near side of that shell, we will see a bias in radial velocity. We
represent this as an average peculiar motion $\vmas\mathbf{e_{*}}$,
where $\mathbf{e_{*}}$ is the unit vector from the Sun to the maser.
We assume that the random component of the maser velocities has a
Gaussian distribution of dispersion $\Delta_v$. In total we take the
probability distribution of the velocities of masers to be
\begin{equation}\label{eq:veldist}
  p(\bolv)\rmd^3\bolv = \frac{\rmd^3\bolv}{(2\pi\Delta_v^2)^{3/2}}
  \exp{\left(\frac{-|\bolv-\overline{\bolv}|^2}{2\Delta_v^2}\right)},
\end{equation}
where
\begin{equation}\label{eq:vbar}
  \overline{\bolv}= -v_c(R)\mathbf{e_\phi}-\vsfr+\vmas\mathbf{e_{*}}.
\end{equation}
Notice that positive values of $v_{\phi,\rm SFR}$ imply that the
masing stars lead Galactic rotation, and that motion of the masers
towards us will be reflected in negative values of $\vmas$. In most
cases we fix both $\vsfr=0$ and $\vmas=0$. The various velocities are 
illustrated in Fig.~\ref{fig:whatvels}.

We sometimes take the Sun's motion with respect to the LSR, $\vsol$,
to be specified, and sometimes we fit it to the data.

For given values of $l_i$, $b_i$, the heliocentric distance $s_i$, and
$R_0$, we deduce the probability distributions in proper motion and
line-of-sight velocity given the above probability distribution of the
velocity (eq.~\ref{eq:veldist}). Since our focus is on
the Galaxy's parameters, we marginalise over the $s_i$.  We do this
under the assumption that the probability distribution of the $s_i$ is
that implied by the assumption of Gaussian errors in the parallaxes.

\subsection{Statistical analysis}

\subsubsection{Likelihood function}
In order to determine the best-fitting model parameters, we maximise
the likelihood function. This is
\begin{equation}
  \mathcal{L}(\bolth) \propto 
  \prod_i \int\rmd s_i\, p(\mathrm{data}\,|\,\bolth),
\end{equation}
where $p(\mathrm{data}\, |\,\bolth )$ is the conditional probability
of the $i$th observation, given the model with parameters
\begin{equation}
  \bolth\equiv( v_0, R_0,\alpha,\vsol,\Delta_v,s_i,\ldots).
\end{equation}
For each distance $s_i$, the model provides a multivariate Gaussian
probability distribution in $v_{r,i}$, $\mu_{\alpha,i}$ and
$\mu_{\delta,i}$.  The data also provide a multivariate Gaussian
distribution for the underlying values of these variables. We obtain
$p(\mathrm{data}\, |\,\bolth )$ by integrating the product of these
two Gaussian distributions over all three variables.

The likelihood function defines the a posteriori probability
distribution of the model parameters $\bolth$. We use a Metropolis
algorithm \citep{Metropolisetal1953} to identify the peak in this
probability distribution, and to characterise its width around the
most probable model. The Metroplis algorithm is a Markov Chain Monte
Carlo method for drawing a representative sample from a probability
distribution, such as the likelihood function. We start with some
choice for the parameters $\bolth$, and calculate the associated
likelihood $\mathcal{L}(\bolth)$. We then
\begin{enumerate}
\item choose a trial parameter set $\bolth '$ by moving from $\bolth$
  in all directions in parameter space, by an amount chosen at random
  (from a Gaussian distribution);
\item determine $\mathcal{L}(\bolth ')$;
\item choose a random variable $r$ from a uniform distribution in the
  range [0,1];
\item if $\mathcal{L}(\bolth ')/\mathcal{L}(\bolth) > r$, accept the
  trial parameter set, and set $\bolth = \bolth '$. Otherwise do not
  accept it.
\item Return to step (i).
\end{enumerate}
The first few values of $\bolth$ are ignored as ``burn-in'', which
helps to remove the dependence on the initial value of $\bolth$.  We
repeat the procedure until the chain of $\bolth$ values constitutes a
fair sample of the probability distribution -- we establish that the
burn-in period is sufficiently long by comparing chains that have
different starting $\bolth$ \citep[e.g.][]{GelmanRubin1992}.

\subsubsection{Bayesian evidence}\label{sec:methods:evid}

In addition to comparing models that differ only in the values taken
by a given set of parameters, we have to assess the value of adding a
parameter to a model. Adding a parameter is guaranteed to increase the
maximum likelihood achievable for given data, but is that increase
statistically significant or just the consequence of an enhanced
ability to fit noise? The Bayesian methodology for making such
assessments is now well established and described by
\cite{Heavens2009}, for example.

One calculates the ``evidence'' for the model under different priors
for whatever parameters are either fixed a priori or varied. The
evidence is the total probability of the data after integrating over
all parameters:
\begin{equation}\label{eq:evid}
  p({\rm data} |{\rm Model}) = \int\rmd^n\bolth\, p({\rm data} | \bolth ,{\rm
    Model})\, p(\bolth |{\rm Model}). 
\end{equation}
Here $p(\bolth |{\rm Model})$ is the prior on the parameters. For a
parameter $\theta_i$ that is varied, we take the prior to be uniform
within a range of width $\Delta\theta_i$ that is large enough to
encompass any plausible value of $\theta_i$ (so $p(\theta_i|{\rm
  Model}) \rmd\theta_i = d\theta_i/\Delta\theta_i$ over this
range). The priors on parameters that are always varied play little
role because we are interested the ratio of the evidence when a
parameter $\theta_n$ is fixed to when it is varied. When $\theta_n$ is
fixed, its prior is a delta function at the chosen value, so the ratio
of the evidences when $\theta_n$ is fixed to when it is varied is
\begin{equation}\label{eq:Eratio} {p({\rm data} |{\rm
      Model},\theta_n\,\fixed)\over p({\rm data} |{\rm Model})} =
  {\int\rmd^{n-1}\bolth\, p({\rm data} | \bolth ,{\rm Model})\over
    \int\rmd^n\bolth\, p({\rm data} | \bolth ,{\rm Model})}\Delta\theta_n,
\end{equation}
where upper integral excludes $\theta_n$ and the lower one includes
it.

By Bayes theorem the ratio of the a posteriori probabilities of the
model when $\theta_n$ is fixed to when it is varied is the ratio of
the corresponding evidences times the ratios of the prior on
$\theta_n$ being fixed to that on it being variable. We take this
second factor to be unity, so the a posteriori probabilities of the
models is simply the ratio of the evidences.

The integration over the space of parameters that is required to
calculate the evidence is exceedingly costly if done by brute
force. We approximate it by assuming that in the vicinity of its peak,
$p({\rm data}|\bolth,{\rm model})$ can be approximated by a Gaussian
$p\propto\exp(-\bolth^T\cdot K\cdot\bolth)$, where $K$ is a matrix
whose eigenvectors and eigenvalues can be estimated from the output of
the Metropolis algorithm. With this approximation, the integral over
parameters becomes analytic.

The Metropolis algorithm yields a set of points in parameter space
which sample $p({\rm data}|{\rm Model})$. Principal component analysis
of this sample yields the eigenvectors and eigenvalues of $K$.

\section{Results}\label{sec:results}

\begin{table*}
  \caption{Log likelihoods for best-fitting models with 
    flat rotation curves ($\alpha=0$),
    power law rotation curves ($\alpha\neq0$) and 
    rotation curves taken from GDII. The first 9 
    likelihoods are for models in which the sources are assumed to have no typical
    velocity --  for the first 3 the value of $\vsol$ is that found in 
    DB98; the next 3 are for models in which 
    $\vsol$ is assumed to be that 
    suggested by B09; and the next 3 are for models in which 
    $\vsol$ is allowed to vary freely; the final 3 are for cases in which 
    $\vsol$ is taken to be the DB98
    value, and the mean source peculiar motion with respect to their local
    standard of rest, $\vsfr$, is allowed to vary freely.} 
  \label{tab:Like}
  \begin{tabular}{c|ccccccccc} 

    & $v_0$ & $R_0$ & $\alpha$ & $U_\odot$ & $V_\odot$ & $W_\odot$ & $\Delta_v$ 
    & $v_0/R_0$ & $\log(\mathcal{L})$ \\
    & $(\mathrm{km\, s}^{-1})$ & $(\mathrm{kpc})$ & 
    & $(\mathrm{km\, s}^{-1})$ & $(\mathrm{km\, s}^{-1})$  
    & $(\mathrm{km\, s}^{-1})$ & $(\mathrm{km\, s}^{-1})$ 
    & $(\mathrm{km\, s}^{-1}\mathrm{kpc}^{-1})$ 
    &  \\ \hline\hline

    \multirow{3}{*}{\rotatebox{90}{DB98}}
    &$200\pm20$ & $6.7\pm0.5$ & $0$ & $10.0$ & $5.2$ & $7.2$ 
    & $10.0\pm1.3$ & $30.1\pm1.7$ & $26.8$\\ 
    &$209\pm26$ & $6.9\pm0.7$ & $0.10\pm0.16$ & $10.0$ & $5.2$ & $7.2$ 
    & $10.0\pm1.3$ & $30.4\pm1.7$ & $27.1$\\ 
    &$218\pm20$ & $7.1\pm0.5$ &GDII & $10.0$ & $5.2$ & $7.2$ 
    & $9.3\pm1.2$ & $30.8\pm1.6$ & $30.7$\\ \hline
    \multirow{3}{*}{\rotatebox{90}{B09}}
    &$215\pm19$ & $7.0\pm0.5$ & $0$ & $10.0$ & $11.0$ & $7.2$ 
    & $8.1\pm1.1$ & $30.5\pm1.5$ & $35.7$\\ 
    &$228\pm24$ & $7.4\pm0.6$ & $0.15\pm0.13$ & $10.0$ & $11.0$ & $7.2$ 
    & $8.0\pm1.1$ & $30.8\pm1.5$ & $36.5$\\ 
    &$235\pm19$ & $7.6\pm0.5$ & GDII & $10.0$ & $11.0$ & $7.2$ 
    & $7.6\pm1.0$ & $31.1\pm1.5$ & $38.9$\\  \hline
    \multirow{3}{*}{\rotatebox{90}{$\vsol$} \rotatebox{90}{free}}
    &$232\pm24$ & $7.7\pm0.6$ & $0$ & $8.1\pm2.9$ & $18.6\pm2.4$ 
    & $9.7\pm2.0$ & $7.1\pm1.0$ & $30.0\pm1.8$ & $42.5$\\ 
    &$258\pm32$ & $8.6\pm0.9$ & $0.27\pm0.13$ & $8.1\pm2.8$ & $19.5\pm2.5$ 
    & $10.1\pm2.0$ & $7.1\pm1.0$ & $29.9\pm1.7$ & $45.4$\\ 
    &$246\pm24$ & $8.1\pm0.6$ & GDII & $8.3\pm2.8$ & $16.5\pm2.4$ 
    & $9.9\pm2.0$ & $7.0\pm1.0$ & $30.3\pm1.8$ & $43.5$\\ \hline\hline

    &$v_0$ & $R_0$ & $\alpha$ & $v_{R,\mathrm{SFR}}$ & $v_{\phi,\mathrm{SFR}}$ 
    & $v_{z,\mathrm{SFR}}$ 
    & $\Delta_v$ & $v_0/R_0$ & $\log(\mathcal{L})$ \\ \hline\hline
    \multirow{3}{*}{\rotatebox{90}{$\vsfr$} \rotatebox{90}{free}}
    &$241\pm24$ & $7.7\pm0.6$ & $0$ & $3.3\pm2.8$  & $-12.9\pm2.4$ 
    & $2.5\pm2.0$ & $7.0\pm1.0$ & $31.1\pm1.7$ & $42.9$\\ 
    &$279\pm33$ & $8.9\pm0.9$ & $0.25\pm0.12$ & $2.0\pm2.8$ & $-14.8\pm2.5$ 
    & $3.0\pm2.0$ & $6.6\pm1.0$ & $31.5\pm1.6$ & $45.5$\\ 
    &$259\pm23$ & $8.2\pm0.6$ & GDII& $2.5\pm2.8$ & $-11.0\pm2.4$ 
    & $2.8\pm2.0$ & $7.1\pm1.0$ & $31.6\pm1.7$ & $43.8$\\ 
  \end{tabular}
\end{table*}

\begin{figure}
  \centerline{\resizebox{\hsize}{!}{\includegraphics{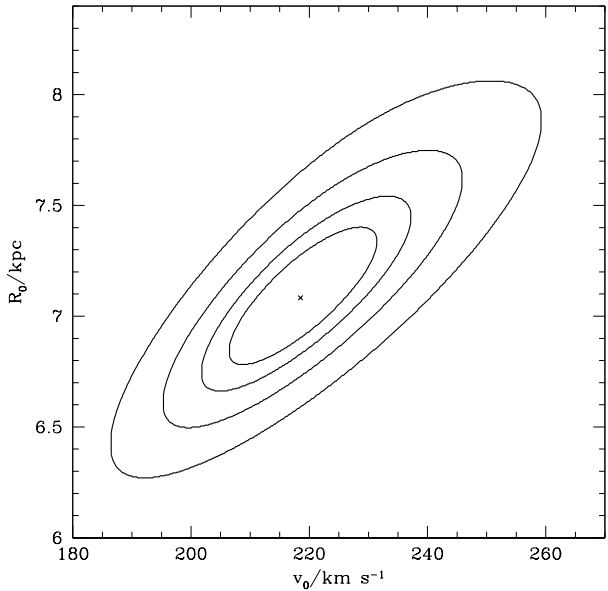}}}
  \caption{Plot showing contours of the Likelihood function (marginalised
    over $\Delta_v$) for a model with the GDII rotation curve, and the 
    DB98 $\vsol$. There is clearly a strong correlation between the
    values found for $v_0$ and $R_0$. Contours are drawn at likelihood 
    differing from the maximum likelihood $\Delta\mathcal{L} =\,0.25$, $0.5$,
    $1$, $2$.
\label{fig:banana}
}
\end{figure}

\begin{figure*}
  \centerline{\hfil
    \resizebox{82mm}{!}{\includegraphics{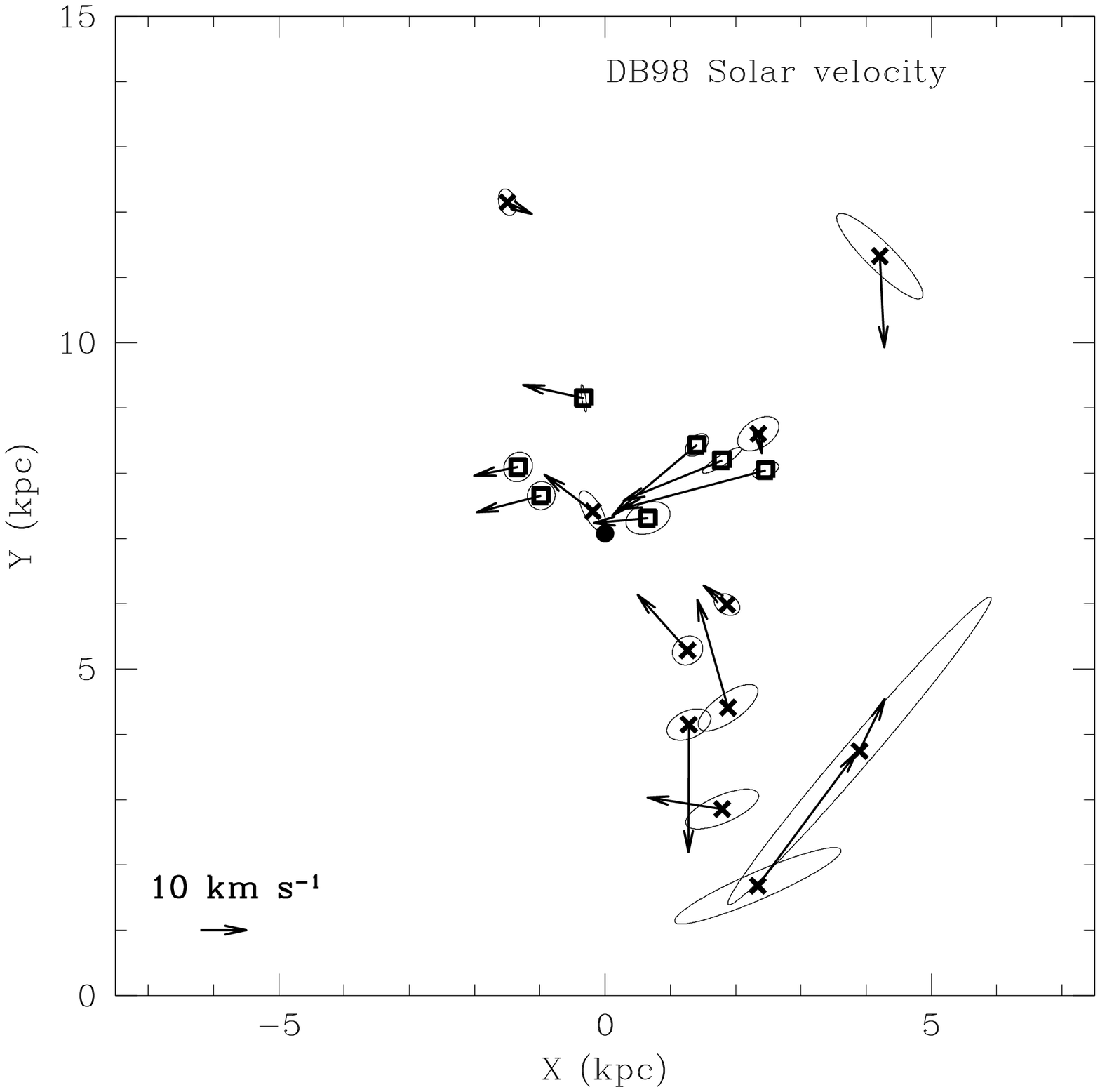}}\hspace{4mm}
    \resizebox{82mm}{!}{\includegraphics{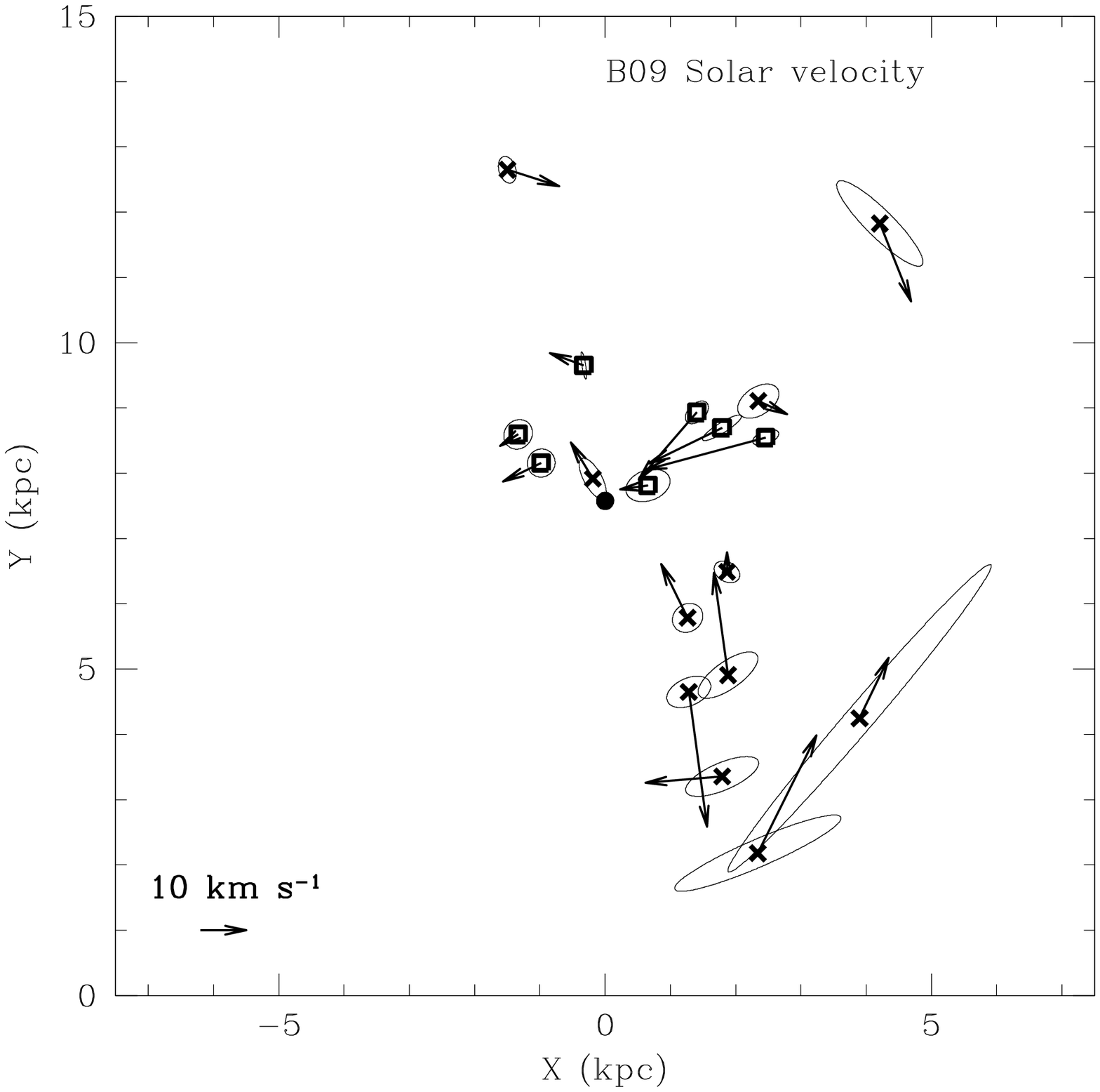}}} \vspace{2mm}
  \centerline{\hfil
    \resizebox{82mm}{!}{\includegraphics{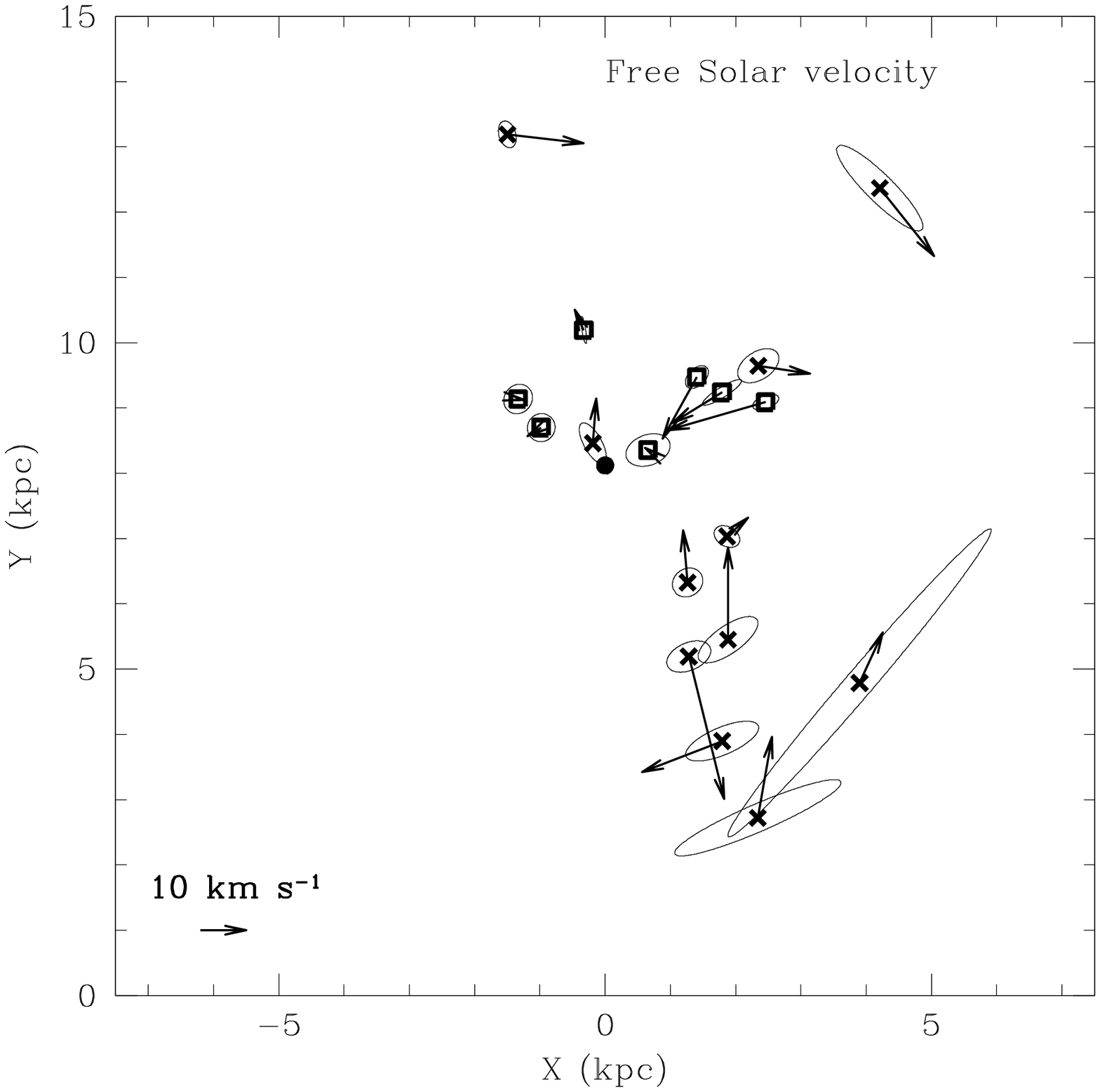}}\hspace{4mm}
    \resizebox{82mm}{!}{\includegraphics{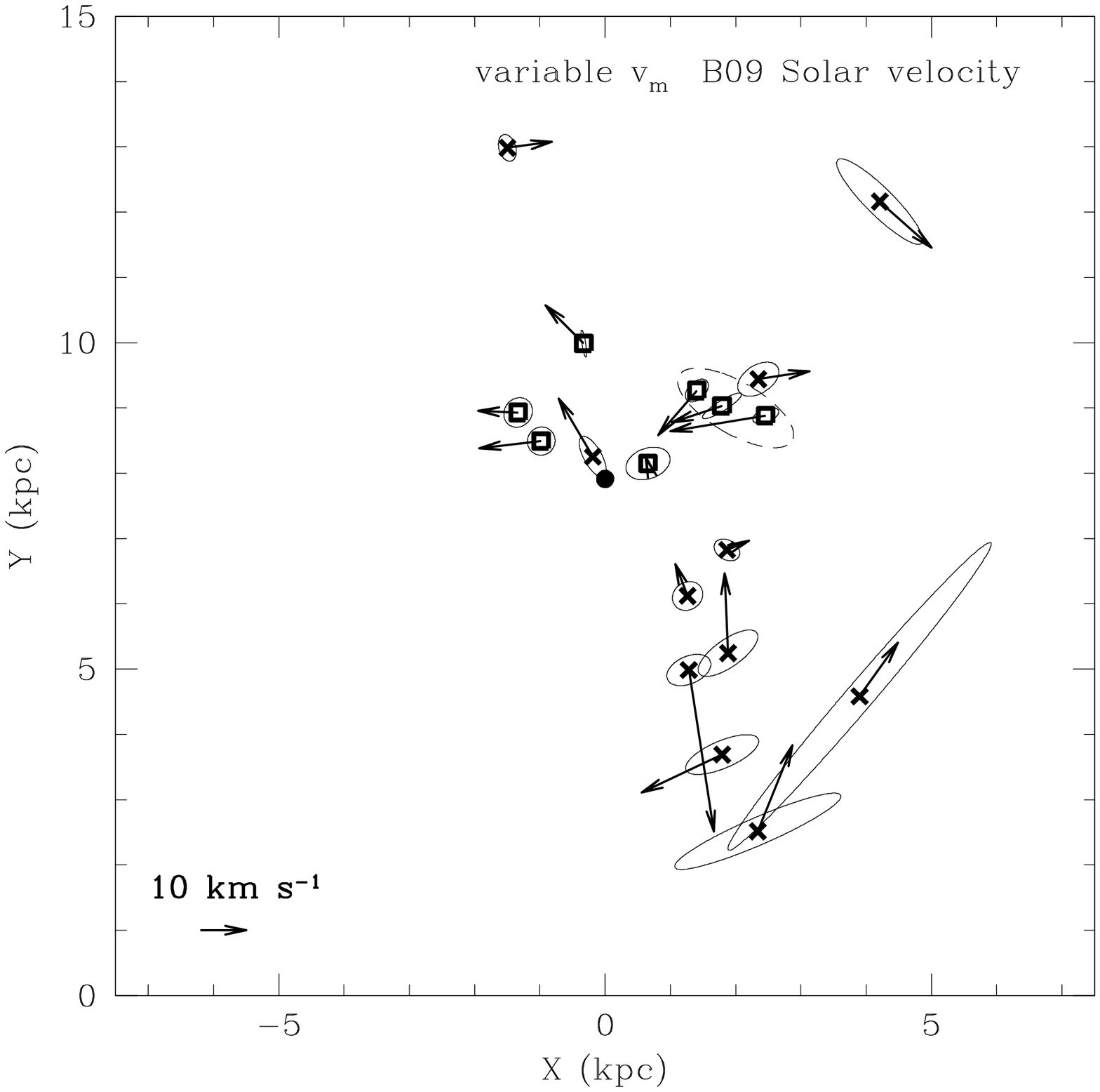}} }
  \caption{
    Residual velocities left after the best-fitting model velocities 
    are subtracted, for models with GDII rotation curves and 
    $\vsol$ taking the DB98 value (top-left); the B09
    value (top-right); or taken to be a free parameter (bottom-left). 
    We also plot the residual velocities left after subtracting the best 
    fitting model velocities for a model with the B09 $\vsol$ value, and
    $\vmas$ taken to be a free parameter (bottom-right). \emph{Open squares} 
    correspond to the objects for which the likelihood improves
    most significantly between the best-fitting DB98 model and the 
    best-fitting free Solar velocity one ($\Delta(\log\mathcal{L}_i)>1$), 
    and \emph{crosses} to all other sources. The Sun is represented by a solid circle. 
    The \emph{solid lined ellipses} around each point are $1\,\sigma$ measurement error 
    ellipses, where the velocity error perpendicular to the line of sight
    in each case is due to a combination of the uncertainty in the proper
    motion and in the parallax.
    The uncertainty in the parallax also means that the residual velocities 
    shown here are not an ideal 
    illustration of the difference between the model and the data, 
    because the peculiar motions can only be given for a chosen
    position (in this case the position corresponding to the quoted parallax).
    However 
    this does provide a useful guide, especially for sources relatively close
    to the Sun, which have small position uncertainties. The \emph{dashed ellipse}
    in the bottom-right plot is drawn around the three objects for which the
    likelihood improves most significantly when $\vmas$ is taken to be a free 
    parameter ($\Delta(\log\mathcal{L}_i)>1$ for B09 $\vsol$) 
   \label{fig:resid}
 }
\end{figure*}

We first investigate models in which the masers are at rest with
respect to their host stars (i.e., $v_{\rm m}=0$).

Table~\ref{tab:Like} gives the peak likelihoods of our models, as well
as the best-fitting parameters and the corresponding
uncertainties. The best-fitting value of $v_0$ varies in the range
$(200-279)\kms$, and that of $R_0$ between $6.7$ and $8.9\kpc$. As
Fig.~\ref{fig:banana} illustrates, $v_0$ and $R_0$ are strongly
correlated with the result that $v_0/R_0$ is confined to the
relatively narrow range $29.9-31.6\kms\,\kpc^{-1}$.

When $\vsol$ is a free parameter, the best-fitting values of $U_\odot$
and $W_\odot$ are close to the DB98 values, and essentially
independent of the form of the rotation curve, whereas $V_\odot$
varies in the range $16.5-19.5\kms$. Similarly, when $\vsol$ is fixed
at the DB98 value and $\vsfr$ is a free parameter, only $v_{\phi,\rm
  SFR}$ takes a value that is far removed from that which one would
naively expect -- it moves in the range $-11.0$ to $-14.8\kms$. Thus
the data suggest either that the Sun is circulating significantly
faster than the circular speed, or that the \HMSFR\ have an
appreciable rotational lag.

The fits are almost perfectly degenerate between $\vsol$ and $\vsfr$:
they constrain only the difference between these velocities. However,
we shall argue in Section~\ref{sec:discuss} that significantly
non-zero values of $\vsfr$ are physically implausible, so here we
focus on what can be inferred about $\vsol$ given $\vsfr=0$. In view
of the degeneracy between $\vsol$ and $\vsfr$ we do not report results
obtained when both $\vsol$ and $\vsfr$ are varied.

When $\vsfr=0$, the peak likelihood is higher when $\vsol$ is fixed to
the B09 value than when it is fixed to the DB98 value. To assess the
significance of this increase in likelihood, we calculate the ratio of
the evidences for the two models as described in
Section~\ref{sec:methods:evid}
\begin{equation} \label{eq:PDB98B09}
  \frac{p(\mathrm{DB98} | \data)} {p(\mathrm{B09} | \data)} \simeq
  \left\{
    \begin{array}{ll}
      2\times10^{-4} & \;\mbox{for }\alpha=0 \\
      1\times10^{-4} & \;\;\alpha\mbox{ variable} \\
      4\times10^{-4} & \;\;\mbox{GDII}. \\
    \end{array}
  \right.
\end{equation}
Thus regardless of the adopted rotation curve, the data strongly
favour upward revision of $V_\odot$ from $5.2\kms$ to $11\kms$.

When $\vsol$ is a free parameter, the peak likelihood of the B09 model
is surpassed at yet larger values of $V_\odot$. To determine whether
the increase in likelihood that occurs when $\vsol$ is set free from
the B09 value, we again calculate the relevant ratio of the
evidences. Since the two models now differ in whether $\vsol$ is fixed
or free, the priors on the components of $\vsol$ now become relevant
(cf eq.~\ref{eq:Eratio}). We have adopted $\Delta U_\odot=\Delta
V_\odot=\Delta W_\odot=100\kms$.  With these values we have
\begin{equation}\label{eq:PB09}
  \frac{p(\mathrm{B09} | \data)} {p(\vsol \vary | \data)} \simeq
  \left\{
    \begin{array}{ll} 
      5 &  \;\mbox{for }\alpha=0 \\
      0.6 & \;\;\alpha\mbox{ variable} \\
      40 & \;\;\mbox{GDII}. \\
    \end{array}
  \right.
\end{equation}
Therefore under these assumptions the increase in likelihood attained
on setting $\vsol$ free from the B09 value is not statistically
significant. The ratio of evidences for the case when $\vsol$ is set
to the DB98 value or is set free is
\begin{equation} \label{eq:PDB98} \frac{p(\mathrm{DB98} |
    \data)}{p(\vsol\vary | \data)} \simeq
  \left\{
    \begin{array}{ll} 
      8\times10^{-4} & \;\mbox{for }\alpha=0 \\
      7\times10^{-5} & \;\;\alpha\mbox{ variable} \\
      0.01 & \;\;\mbox{GDII}. \\
    \end{array}
  \right.
\end{equation}
Therefore, even with a generous choice of $\Delta U_\odot$, etc., the
data reject the possibility that the Sun has the DB98 value of
$\vsol$. Reducing the widths $\Delta U_\odot$, etc., of the priors on
$\vsol$ would strengthen the case against the DB98 value of $\vsol$.

The choice of $\Delta U_\odot \sim 100\kms$ is reasonably generous,
and it is sensible to ask what value of $\Delta U_\odot$, etc., would
bring ${p( \mathrm{B09} | \data)}/ {p(\vsol\vary | \data)}$ down to
unity. In the case where the GDII rotation curve is used, it would
require a reduction to a value $\Delta U_\odot \sim 30\kms$ -- this is
smaller than is reasonable given that the value for $V_\odot$ found
when it is allowed to vary is already $\sim15\kms$ greater than the
canonical DB98 value.

It is worth noting that the case for setting $\vsol$ free from either
the DB98 value of the B09 value is weakest when the best-motivated
rotation curve is adopted -- the GDII curve.

\begin{table*}
  \begin{tabular}{ccccccccc} 

    $v_0(\kms)$ & $R_0(\kpc)$ & $U_\odot$ & $V_\odot$  & $W_\odot$ 
    &$\vmas$ & $\Delta_v$ & $v_0/R_0(\kms\,\kpc^{-1})$ & $\log(\mathcal{L})$ \\ \hline
    $233\pm20$ & $7.3\pm0.5$ & $10.0$ & $5.2$ & $7.2$ & $-8.1\pm2.7$ & $8.0\pm1.1$ & $31.6\pm1.5$ & $36.2$\\
    $247\pm19$ & $7.8\pm0.4$ & $10.0$ & $11.0$ & $7.2$ & $-6.2\pm2.4$ & $6.8\pm0.9$ & $31.4\pm1.4$ & $43.1$\\
    $259\pm25$ & $8.2\pm0.5$ & $10.7\pm3.3$ & $15.0\pm2.7$ & $10.0\pm2.0$ & $-4.8\pm2.9$ & $6.5\pm1.0$ & $31.4\pm1.9$ & $46.0$\\ \hline\hline
    $v_0(\kms)$ & $R_0(\kpc)$ & $v_{R,\mathrm{SFR}}$ & $v_{\phi,\mathrm{SFR}}$  
    & $v_{z,\mathrm{SFR}}$ & $\vmas$ & $\Delta_v$ & $v_0/R_0(\kms\,\kpc^{-1})$ 
    & $\log(\mathcal{L})$ \\ \hline\hline
    $271\pm25$ & $8.2\pm0.6$ & $2.0\pm3.1$ & $-9.7\pm2.8$ & $-2.5\pm2.2$ 
    & $-5.8\pm3.3$ & $6.4\pm1.1$ & $32.9\pm2.0$ & $46.0$\\ 

  \end{tabular}
  \caption{
    Similar to Table~\ref{tab:Like}. This shows the best-fitting parameters (and
    corresponding log likelihoods) for models with a GDII rotation curve, and
    with the value $\vmas$ allowed to vary.  
  }
  \label{tab:maser}
\end{table*}

The lowest values of both $v_0$ and $R_0$ are found for the models
with the DB98 value of $\vsol$, and $\vsfr$ set to zero. Both $v_0$
and $R_0$ take larger values for models with the B09 value of $\vsol$,
and larger still for models with either $\vsol$ or $\vsfr$ allowed to
vary. The lowest values of $v_0$ and $R_0$ for the models in which the
peculiar velocities are allowed to vary are $232\kms$ and $7.7\kpc$
respectively.

Fig.~\ref{fig:resid} suggests a reason why the value of $\vsol$ or
$\vsfr$ in the model has such a large impact on the best fitting values of 
$v_0$ and $R_0$.  It
shows the residual velocities of the objects after the expected
velocity is subtracted (ignoring the uncertainty in parallax). The
objects that provide the strongest indication that some change in
peculiar motion is required (indicated by open squares)
are all relatively close to the
Sun. Consequently, changes in $R_0$ and $v_0$ have a relatively small
``lever arm'' with which to bring the model closer to the data, with
rather large changes required in order to have any significant impact
on the expected values of the observables for those objects.  
The introduction of a new
peculiar velocity (either of the sources or the Sun) allows the model
to compensate for the significant discrepancy between the velocity of
these objects and those expected from the rotation curve, without
producing a major effect on the best-fitting values of $R_0$ and
$v_0$.

\subsection{Maser line-of-sight velocities}

Finally we consider models in which we allow for a systematic
difference between the measured line-of-sight velocity of a maser and
the line-of-sight velocity of the exciting star -- such a difference
would arise if masing occurred on the near side of an expanding shell
around the star. We have done this for all the rotation curves
described in this paper, but the results are sufficiently similar to
one another that we only present (in Table~\ref{tab:maser}) the
results for the GDII rotation curve.

We again compare the different models by calculating the ratios of
their evidences, using the prior for $\vmas$ uniform in a range of
width $\Delta\vmas = 20\kms$.  We find
\begin{equation}\label{eq:usevm}
  \frac{p(\vmas = 0 | \data)} {p(\vmas\vary | \data)} \simeq
  \;\left\{
    \begin{array}{ll} 
      0.01 & \;\mbox{for DB98} \\
      0.06 & \;\; \mbox{B09} \\
      0.3 & \;\;\vsol\vary \\
      0.3 & \;\;\vsfr\vary \\
    \end{array}
  \right.
\end{equation}
so in each case the data support adding the extra free parameter.

If we accept this extra parameter in our models, we have to reconsider
the evidence for a change in $\vsol$ (or $\vsfr$). The relevant ratios
of evidences are:
\begin{equation} 
  \frac{p(\mathrm{DB98}, \vmas\vary | \data)}{p(\mathrm{B09}, \vmas\vary | \data)} \simeq
  0.001,
\end{equation}
\begin{equation}
  \frac{p(\mathrm{DB98}, \vmas\vary | \data)} {p(\vsol , \vmas\vary | \data)} \simeq
  0.3,
\end{equation}
\begin{equation}
  \frac{p(\mathrm{B09}, \vmas\vary | \data)}{p(\vsol , \vmas\vary | \data)} \simeq
  220.
\end{equation}
So the B09 value of $\vsol$ is still favoured. Moreover, when $\vsol$
is taken to be free, the likelihood still peaks at an even larger
value of $V$ than that of B09.

We note that the sources providing the strongest evidence that a
non-zero value of $\vmas$ is needed are those ringed in the the
bottom right plot of Fig.~\ref{fig:resid}, all associated with the
Perseus spiral arm \citep{Reidetal2009}. If we exclude these sources,
the best fitting value of $\vmas$ is $\sim -3\kms$ for any assumption
about $\vsol$ we consider.  This is approximately within the
uncertainty on $\vmas$, and if we considered only this subset of the
data in equation~(\ref{eq:usevm}), the evidence would not support adding
the extra free parameter $\vmas$ (though it should be noted that it is
still a radial velocity \emph{towards} the Sun).  It is, therefore,
possible that what we have modelled as an offset in the radial
velocities of all observations is actually primarily due to a large
peculiar velocity of the objects in the Perseus arm, directed
approximately in the direction of the Sun.

\section{Discussion}\label{sec:discuss}

These results, and in particular the large variation in $v_0$ and
$R_0$ depending on the other model parameters, makes it impossible to
constrain tightly either $v_0$ or $R_0$ from these data -- the
smallest best-fitting values of these parameters are, respectively, 40
per cent and 33 per cent smaller than the largest best-fitting values.
The choice of rotation curve has a significant impact, and we do not
investigate all possible rotation curves. It is worth noting that our
two most favoured models in Table~\ref{tab:Like} (according to Bayesian 
evidence, so somewhat dependent on our choice of priors) have best-fitting 
values of $v_0$ that differ by $30\kms$ and of $R_0$ that differ by
$1\kpc$ from one another
(the models are those with B09 $\vsol$ $+$ GD rotation curve and free 
$\vsol$ $+$ power-law rotation curve, respectively).

However, for all but one of our models the best fitting $v_0/R_0$ lie
in the narrow range $(29.8-31.5)\kms\kpc^{-1}$, with typical
uncertainties $\sim1.5\kms\kpc^{-1}$. This corresponds to best
fitting $(v_0+V_\odot)/R_0$ in the range $(30.9-32.5)\kms\,\kpc^{-1}$
(again, depending on the model), with similar uncertainties.  These
values are slightly larger than the value of $(v_0+V_\odot)/R_0 =
30.2\pm0.2\kms\kpc^{-1}$ found from the proper motion of Sgr
A*~\citep{ReidBrunthaler2004}, but consistent to within (less than) twice
the uncertainties on our values.

If we allow for a bias in the maser radial velocities with respect to
the Sun by allowing the parameter $\vmas$ to take a non-zero value in
equation~(\ref{eq:vbar}), this does improve the fit. However we have
seen that this is primarily due to a group of maser sources in the
Perseus spiral arm, so the cause may be a large peculiar motion in
that part of the Galaxy. Even if the bias in measured radial velocity
is real, it does not affect the result: (i) that in different models these
data are fit best by very different values of $v_0$ and $R_0$; (ii) that
the DB98 $\vsol$ is rejected by these data, and (iii) that the B09 $\vsol$
is favoured over making $\vsol$ a free parameter.

\cite{Reidetal2009} focused on the idea that these data show that the
\HMSFR s are orbiting the Galaxy with a velocity that lags the
circular speed by $-\vpsfr\sim15\kms$. Our analysis of similar models
shows that the data are fit better by a slightly smaller offset of
$(11-14.8)\kms$ to the circular speed. We have also shown that an
equivalent fit to the data can be obtained by instead increasing to
$\sim17\kms$ the amount $V_\odot$ by which the Sun is assumed to
circulate faster than the circular speed.  In fact, the sources that
provide the strongest statistical support for $\vpsfr\sim-15\kms$ are
found near the Sun (Fig.~\ref{fig:resid}), so a change in $\vsol$ has
a very similar effect to a change in $\vsfr$. More maser data at
different Galactic azimuths could break this degeneracy, as well as
reducing the correlation between the values of $v_0$ and $R_0$ seen in
Fig.~\ref{fig:banana}.

If the proposal of \cite{Reidetal2009} that $\vpsfr\simeq-15\kms$ were
correct, the \HMSFR s would all have to be close to apocentre. Since
the lifetime of an \HMSFR\ is short compared to a typical epicycle
period and spiral structure must play a significant role in their
formation, it is not inherently implausible that the maser stars are
all close to apocentre. However, two arguments make it unlikely that
the maser stars are on orbits as eccentric as is implied by a $15\kms$
offset from circular motion at apocentre.

First, non-axisymmetric structure in the Galaxy's potential modulates
the tangential velocity of gas by only $\sim7\kms$ \citep[e.g.][\S
9.2.3]{GA}, so the \HMSFR s would be moving with a peculiar velocity
significantly higher than the gas from which they formed. Second, by
GDII (equation 3.100), a quarter of an epicycle period later the
$U$ velocities of these stars would be
$(2\Omega/\kappa)15\kms\simeq22\kms$, where $\Omega$ and $\kappa$ are
the circular and radial frequencies at the star's location. Hence the
radial velocity dispersion of a population of such stars would be at
least $\sim22/\surd2\simeq15\kms$.  The velocity dispersion of stars
observed locally increases with age on account of heating by
irregularities in the Galaxy's gravitational field (e.g. GDII \S
10.4.1) and the velocity dispersion of the bluer stars in the
Hipparcos catalogue is $\sim10\kms$ \citep{AumerBinney2009} rather
than $15\kms$.  Thus the conjecture of \cite{Reidetal2009} not only
requires the masing stars to be confined to apocentre but also
requires them to have more eccentric orbits than the generality of
young stars. Recognising this problem, Reid et al.\ suggested that the
orbits of these stars became more circular early in their
lives. However, any random scattering process spreads stars more
widely in phase space and therefore increase the mean eccentricity of
the masing stars. Only a dissipative process could increase the
phase-space density of these stars by moving them to more circular
orbits, and no such process is known.

If the B09 value of $V_\odot$ is correct, the lag to circular motion
required to optimise the fit to the data is so small ($\sim5\kms$)
that the above objections to the proposal that \HMSFR\ systematically
lag rotation become moot. However, the formalism of Bayesian inference
says that when the B09 value of $V_\odot$ is accepted, there is no
convincing evidence for a systematic lag of the \HMSFR.

Throughout this paper we have proceeded under the assumption that 
the velocity of the LSR is the same thing as the 
circular velocity at $R_0$. It is 
worth noting that this is not necessarily the case. The LSR is defined 
to be the velocity of a closed orbit as it passes the current Solar 
position, which will only be a circular orbit if the Galactic potential 
is axisymmetric. Therefore the velocity of the LSR may be offset from the 
circular velocity curve. However (much like the possibility of a 
systematic lag of the \HMSFR) we should note that if we 
accept the B09 value of $V_\odot$, there is no convincing evidence for 
this offset.

We conclude that these data provide a compelling case for revising
$V_\odot$ upward.  Our Bayesian analysis (equation~\ref{eq:PB09})
supports the revision upwards of $V_\odot$ to $11\kms$ suggested by
B09 on the basis of modelling predominantly old disc stars with
relatively large random velocities. It should, however, be recognised
that an even better fit to \emph{these} data comes from somewhat
larger upward revisions to $V_\odot \sim 16 - 20\kms$.

\section{Conclusions}

We have reanalysed observations of \HMSFR s reported in
\cite{Reidetal2009} using a maximum-likelihood approach to exploit
fully the data, in an effort to determine the values of $R_0$, the
Galactocentric radius of the Sun, and $v_0$, the circular velocity of
the LSR. We have found that the best-fitting
values, considered separately, are strongly dependent on the Galaxy
model used to interpret the data, but that the ratio $v_0/R_0$ is
consistently found to lie in the range $29.8-31.5\kms\,\kpc^{-1}$.

We have also used these data to explore the value of the Sun's
peculiar velocity $\vsol$, in light of the recent argument of B09 that
the canonical DB98 value is incorrect. We find that these data support
the conclusion that $V_\odot$ is significantly higher than the DB98
value. By a small but significant margin, the data prefer models with
the B09 revision of $V_\odot$ to $\sim 11\kms$ over models in which
$\vsol$ is left as a free parameter (equation~\ref{eq:PB09}).  The
best-fitting models have $U_\odot$ and $W_\odot$ near to the DB98
values, and the even larger value of $V_\odot \sim 16 - 20\kms$.

\cite{Reidetal2009} suggested that \HMSFR s significantly lag circular
rotation. We have investigated the possibility that the \HMSFR s have
a typical peculiar velocity $\vsfr$ and find that the models only
constrain the velocity difference $\vsfr-\vsol$. We have argued that
models in which the \HMSFR s have large peculiar velocities in the
opposite direction to Galactic rotation are neither needed nor
plausible.

This work, in conjunction with B09, casts severe doubt on the accuracy
of the widely used DB98 value for $V_\odot$. This must be a concern
for anyone interested in the dynamics within the Milky Way because
velocities are inevitably measured with respect to the Sun.  Both
observations of more masers and the development of a detailed
dynamical model of the Galaxy's spiral structure would contribute to
establishing more securely what the true value of $V_\odot$ is.

\section*{Acknowledgments}

We thank John Magorrian and the other members of the Oxford dynamics
group for valuable discussions. PJM is supported by a grant from the
Science and Technology Facilities Council.

\section*{Note added after acceptance}

After this paper was accepted for publication, \cite{Rygletal2009} 
presented similar observations of maser sources. Incorporating 
these observations in our analysis does not materially affect our 
conclusions.

\bibliographystyle{mn2e} \bibliography{refs}
\end{document}